\documentclass[%
 reprint,
 amsmath,amssymb,
 aps,
]{revtex4-2}

\usepackage{graphicx}
\usepackage{dcolumn}
\usepackage{bm}
\usepackage[mathlines]{lineno}
\usepackage{bm}
\usepackage[utf8]{inputenc}
\usepackage{amssymb}
\usepackage{color}
\usepackage{amsmath}
\usepackage{lipsum}
\usepackage{subcaption}
\usepackage{caption}
\def\bs{{\textbf{s}}}

\begin{document}

\preprint{APS/123-QED}

\title{Single-molecule Automata: Harnessing Kinetic-Thermodynamic Discrepancy for Temporal Pattern Recognition}

\author{Zhongmin Zhang}
\author{Zhiyue Lu}%
 \email{zhiyuelu@unc.edu}
\affiliation{%
 Department of Chemistry, University of North Carolina at Chapel Hill, Chapel Hill, NC 27599-3290}%

\date{\today}

\begin{abstract}
Molecular-scale computation is crucial for smart materials and nanoscale devices, yet creating single-molecule systems capable of complex computations remains challenging. We present a theoretical framework for a single-molecule computer that performs temporal pattern recognition and complex information processing. Our approach introduces the concept of an energy seascape, extending traditional energy landscapes by incorporating control parameter degrees of freedom. By engineering a kinetic-thermodynamic discrepancy in folding dynamics, we demonstrate that a linear polymer with $N$ binary-state foldable units can function as a deterministic finite automaton, processing $2^N$ configurations. The molecule's dominant configuration evolves deterministically in response to mechanical signals, enabling recognition of complex temporal patterns. This design allows complete state controllability through non-equilibrium driving protocols. Our model opens avenues for molecular-scale computation with applications in biosensing, smart drug delivery, and adaptive materials. We discuss potential experimental realizations using DNA nanotechnology. This work bridges the gap between information processing devices and stochastic molecular systems, paving the way for sophisticated molecular computers rivaling biological systems in complexity and adaptability.
\end{abstract}

\keywords{}
\maketitle

\section{Introduction}
Information processing is fundamental to biological and artificial smart materials, with information encoded in various forms such as DNA sequences \cite{goldman2013towards,tabatabaei2015rewritable,erlich2017dna,organick2018random}, chemical concentrations \cite{qian2011scaling,erbas2018molecular,katz2010enzyme,miyamoto2013synthesizing,cardelli2017efficient}, or molecular configurations \cite{motlagh2014ensemble}. The ability of chemical systems to respond to incoming information is crucial for processes ranging from cellular signaling to the design of smart materials and molecular machines \cite{erbas2015artificial}. Recent advances in chemical computation have led to the development of systems using DNA strand displacement \cite{soloveichik2010dna,qian2011scaling} to perform computational tasks.

While artificial chemical computation systems described above often rely on one-shot processes, living systems demonstrate more complex computational capabilities. Gene regulatory networks and enzyme cascades \cite{purvis2013encoding,bhalla1999emergent,maity2020information} can process and respond to intricate temporal patterns of signals, guiding cellular decisions and maintaining homeostasis. Inspired by these natural systems, researchers have developed artificial biochemical networks capable of processing time-varying input signals \cite{elowitz2000synthetic,bashor2019complex}. However, these approaches often depend on complex networks with numerous interacting components, presenting challenges in design and control, especially at smaller scales.

In this work, we focus on demonstrating continuous-time computation at the individual molecular scale. By harnessing conformational dynamics and the information-processing capabilities of single molecules, we aim to develop a new paradigm for molecular computation that rivals the complexity and adaptability of biological systems. Our approach has the potential to enable the design of smart, responsive materials and devices that can process and respond to real-time information at the nanoscale.

\begin{figure*}
    \centering
    \includegraphics{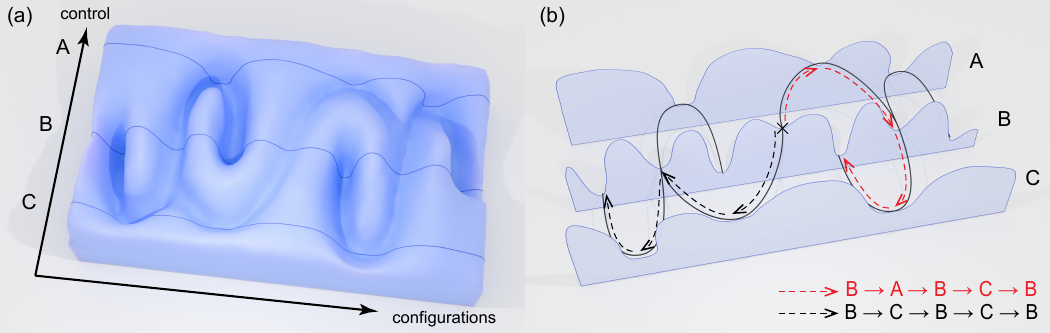}
    \caption{Energy seascape concept and its impact on molecular state transitions. (a) Schematic representation of an energy seascape incorporating both configuration and control parameter degrees of freedom. Three energy landscapes at fixed control conditions A, B, and C are highlighted by light blue curves. (b) Illustration of how different control protocols (sequences of ABC's) can lead to distinct state transition paths (dashed arrows) and ultimate states, even if they start from the same initial state (labeled by a cross.}
    \label{fig:cartoon}
\end{figure*}

\section{Results}
\subsection{Configuration-based Computation on Rugged Energy Seascape}

Here, we demonstrate a single-molecule computation process that is capable of complex tasks such as temporal pattern recognition. The computation is achieved via configuration change of the single molecule in response to the external input information.

A general theory for configuration-based computation is developed based on the concept of \emph{Energy Seascape Design}. Energy seascapes \cite{gaspar1804protein,mustonen2009fitness} extend traditional energy landscapes \cite{dill2012protein,wolynes2015evolution,wales2018exploring} by incorporating control-parameter degrees of freedom alongside configuration degrees of freedom (see Figure~\ref{fig:cartoon}). This key difference enables modeling of system dynamics under varying external conditions, in contrast to traditional landscapes which only describe energetics under fixed conditions. Given an energy seascape, one can find a traditional energy landscape corresponding to each possible value of the control parameter. By varying the control parameter according to a specific time protocol, the corresponding energy landscape ``dances" over time, steering the system's dynamics along different paths. For example, starting from the same state, different input protocols can guide the system into distinct ultimate states, as shown in Fig.~\ref{fig:cartoon}b. The resulting path is determined by the interplay between the system's initial state, energy seascape, and the control protocol. By considering the control protocol as a temporal-sequence-type input signal and the ultimate state of the system as the output, the system can be regarded as performing real-time computation or temporal pattern recognition.

While a rugged energy seascape may allow a system to exhibit complex responses toward temporal signals, it is not sufficient to guarantee computation. In general, complex systems may respond to external controls in various ways, but to qualify as a computation process, the system must exhibit diverse and distinctive output states. For a single-molecule configuration-based computer, this poses a significant challenge. On one hand, the molecule must possess multiple metastable configurations to ensure diverse output states. On the other hand, a single molecule operating in a thermal environment may not distinctively occupy a single configuration but rather spread its probability across many states. 

In this work, we propose a design principle that overcomes this challenge by engineering the discrepancy between the thermodynamic and kinetic trends. This allows a multi-state molecule to perform computation effectively. Furthermore, we will demonstrate that the molecule functions exactly as a deterministic finite automaton, a type of computational model used for language recognition.

\begin{figure*}
    \centering
    \includegraphics{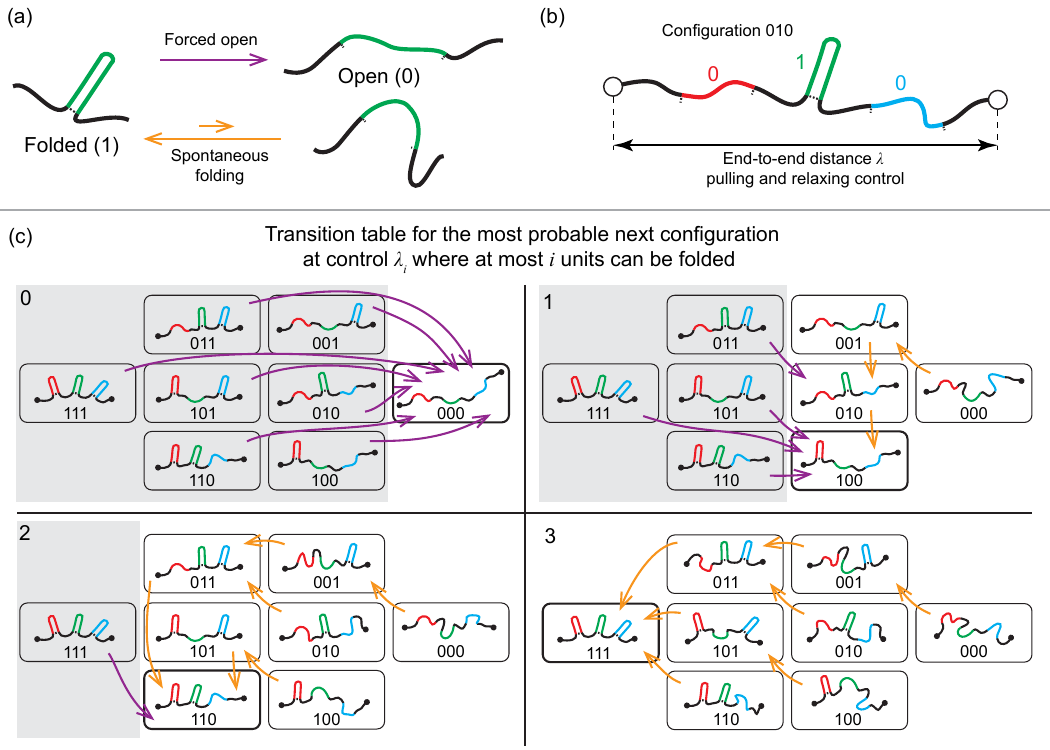}
    \caption{Schematic of the single-molecule automaton and its operation. (a) Representation of one unit of the molecule under transition between folded and unfolded states. (b) Illustration of a 3-unit molecule’s under the mechanical control (in terms of end-to-end distance $\lambda$). (c) State transition tables of the $2^3$-state molecule under different control parameters $\lambda_i$, illustrating how the dominant configuration evolves under various conditions, similar to a deterministic finite automaton.}
    \label{fig:3unit}
\end{figure*}

\subsection{Computation based on dominant configuration time evolution}

We present a proof-of-principle single-molecule computer that operates as a \emph{deterministic finite automaton} \cite{Moore2011-pa,hopcroft2001introduction}, demonstrating how complex computation can be achieved at the molecular scale. This molecule is a linear polymer made by joining $N$ binary-state foldable segments, as a result it consists of $2^N$ distinct configurations. In this linear molecule, each segment (unit) can either be at a folded state ``1" or an open state ``0". Each distinct configuration of the molecule can thus be labeled by $N$-bit string of $1$ and $0$, e.g. ${\bf{s}}=``010"$. An example of such a model molecule with $N=3$ is shown in Fig.~\ref{fig:3unit}a,b). 

Under thermal fluctuation all units in a molecule can fold and unfold stochastically. Without external interference, we assume that each unit tend to fold spontaneously. 
We also assume that each foldable unit can assume two length -- shorter at the folded state and longer at the open state. The total length of the molecule at each given configuration is denoted by the \emph{effective contour length}, $l(\bf{s})$.

The molecule is designed to respond to mechanical signals -- the externally controlled \emph{end-to-end distance} of the molecule, $\lambda$. 
Under a short end-to-end distance, the units in the molecule can spontaneously fold, resulting in a shorter effective contour length. In comparison, under a large end-to-end distance, the molecule's configuration is limited to fewer folded units. Under each end-to-end distance, there is a maximum number of folded units as allowed by the physical restraint: $l\geq \lambda$.
Here we conveniently discretize the mechanical signal $\lambda$ (end-to-end distance) into $N+1$ possible values, $\lambda_i$ for $i=0,1,\cdots,N$, which reflects the maximal number of folded units allowed by the end-to-end distance of the polymer. 
By designing the kinetic and thermodynamic properties of each unit's folding-unfolding dynamics, the molecule's configuration can be steered by the energy seascape with $N+1$ possible transient landscapes to traverse different paths.

As a result, the dominant configuration of the molecule encodes information of the previous temporal patterns of the mechanical control signal, and can be used to perform pattern recognition. 
Even though the dynamics of each individual molecule is stochastic with thermal fluctuations, we find that the \emph{dominant configuration} -- the configuration of maximum probability -- evolves under deterministic rules. (see Fig~\ref{fig:3unit}c)
Moreover, the molecule's dominant configuration dynamics exhibits deterministic-finite-automaton-like behavior as defined in the field of computer science. 

The focus on dominant configurations rather than full probability distributions presents us with both practical and theoretical advantages. In practice, experimentally measurement of the most probable state is far simpler than obtaining a complete distribution, especially for systems with many states. Theoretically, this approach allows us to describe the system's behavior using deterministic rules, effectively bridging the gap between stochastic molecular dynamics and deterministic computation. Specifically the deterministic evolution of the dominant configuration allows us to treat the molecule as a deterministic finite automaton. This simplification is crucial as it allows the dominant configuration to encode information about previous temporal patterns of the mechanical control signal, thereby enabling pattern recognition. By focusing on dominant configurations, we transform an inherently probabilistic molecular system into a reliable computational device that can process and respond to complex temporal inputs.

The molecule's dominant-configuration dynamics as a deterministic finite automaton is described as follows in the language of computer science \cite{Moore2011-pa,hopcroft2001introduction}.  
The set of $N+1$ symbols $\lambda_i$ form an ``alphabet'', $\{\lambda_i\}$. 
The transitions of the machine states (evolution of dominant configurations) at any given symbol $\lambda_i$ (mechanically controlled length) is illustrated by a state transition map (shown in Fig.~\ref{fig:3unit}c for each possible ``symbol'' from the ``alphabet''). This automaton, starting from a given initial state (initial dominant configuration) evolves according to the four transition maps given a temporal sequence of symbols according to the mechanical length protocol, may result in distinct ultimate state (final dominant configuration) that encodes information of the symbol sequence (temporal pattern of the mechanical protocol).
This formulation of deterministic finite automaton provides a powerful framework for designing and analyzing molecular-scale computation.

\subsection {Design principle-- Kinetic-thermodynamic Discrepancy}
The key principle to design a molecular computer is the creation of a discrepancy between thermodynamic stability and kinetic accessibility, which we term \emph{thermal-kinetic frustration}. As a result of this design principle, at any given control, the easiest states for the molecule to reach are not its most stable states, and vice versa.

\begin{figure*}
    \centering
    \includegraphics{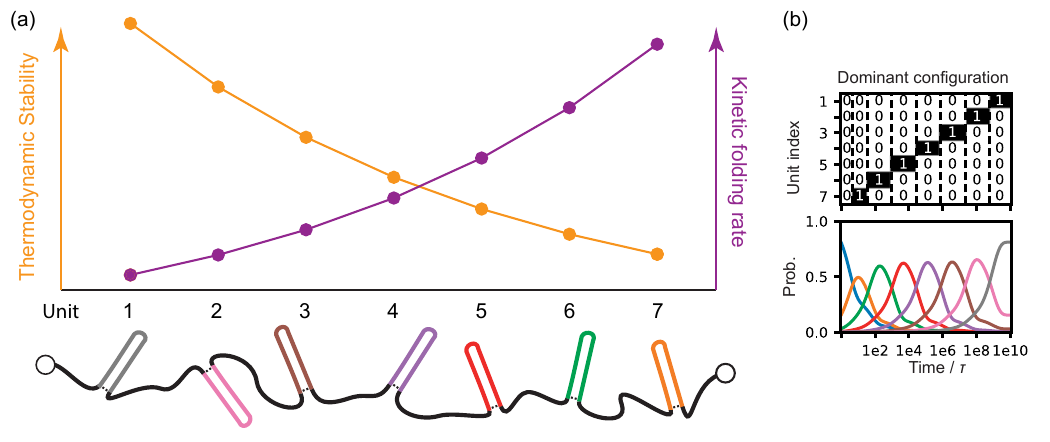}
    \caption{Design principle of thermal-kinetic frustration. (a) Illustration of the discrepancy between thermodynamic stability and kinetic folding rates for seven foldable units. From unit index 1 to 7, stability decreases yet kinetic accessibility (rates) increases. (b) The resulting ``hourglass" behavior of a 7-unit molecule, starting at initial state ``0000000", under a fixed external control allowing up to one folded unit. The top panel illustrates the evolution of dominant configurations, and the bottom panel plots their transient probabilities over time.}
    \label{fig:design}
\end{figure*}

This frustration serves the purpose of delaying equilibrium: By making the thermodynamically stable states kinetically difficult to access, we prevent the molecule from directly settling into its equilibrium configuration. This allows the molecule to traverse a number of meta-stable configurations for an extended period, enabling it to process and respond to time-varying inputs.

To illustrate this concept in our model, we design that, for units indexed from ``1" to ``7", the folded states are less thermodynamically stable for larger index units, however, the kinetic folding rates are larger for larger index, as shown in Fig.~\ref{fig:design}a. During the folding transitions, the least stable unit 7, is most likely to fold first, followed by the more stable units, ordered from 6 to 1. This design leads to the "hourglass" behavior shown in Fig.~\ref{fig:design}b. When we allow the molecule to fold only one unit, it doesn't immediately jump to its most stable configuration (1000000). Instead, it follows a predictable sequence of states, starting with the easiest-to-reach but least stable state (0000001), then progressing through intermediate states (0000010, 0000100, etc.) before finally reaching the most stable state. This design guarantees the dominant configurations to evolve by shifting the position of the folded unit from index 7 to index 1, and their transient probability over the whole time period are shown in Fig.~\ref{fig:design}b. 

The resulting "hourglass" behavior is crucial for enabling complex computation in a single molecule. It ensures a time-dependent response to inputs, allowing the molecule to effectively "remember" the time duration under a fixed control condition. This temporal dimension enables pattern recognition and other sophisticated computational tasks that would be impossible with a simple equilibrium-seeking system.

In summary, our design rule guarantees that at a constant control condition $\lambda$, the molecule traverse a complex path toward equilibrium. As control parameter $\lambda$ changes in time, the accessible state space (i.e., mechanically allowed configurations) is updated as well. As a result, the frustrated energy seascape diversifies the transition paths of the molecule's configurations: It allows the molecule to respond to different control protocols by creating distinct pathways in the configuration space. These separate pathways lead to different ultimate configurations (output), allowing the molecule to effectively classify different input patterns.

\subsection{Logical Evolution Rule for $N$-unit molecules} 
\begin{figure*}[ht!]
    \centering
    \includegraphics{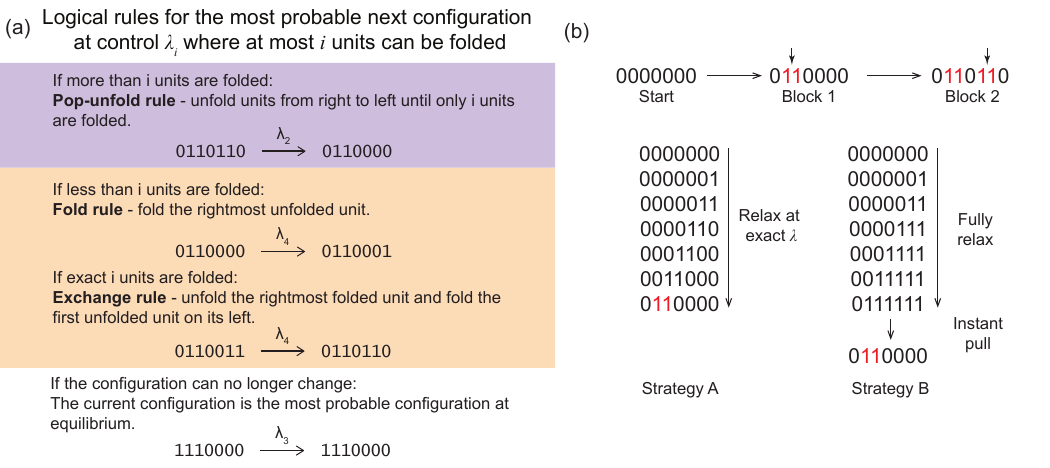}
    \caption{Logical evolution rules and control strategies to steer the molecule's dominant configuration. (a) Flowchart describing the three logical rules governing the evolution of dominant configurations. (b) Comparison of two strategies (A and B) for preparing specific folded configurations, which can be used toward complete state controllability.}
    \label{fig:rules}
\end{figure*}

The dominant configuration dynamics of our designed molecule follows similar deterministic rules for different choices of unit number $N$. This similarity can be illustrated by the hourglass example (Fig.~\ref{fig:design}b) which works similarly for different unit number $N$. However, for large unit number (e.g., $N=7$) the number of configurations is too large to be shown by the configuration transition graph, as in \ref{fig:3unit}c. Here we find that for the designed $N$-unit molecule, the dynamical transitions of dominant configurations has a simple alternative representation in addition to the maps shown in Fig.~\ref{fig:3unit}c.

By observing all dominant configuration dynamics, one can conclude an alternative representation of the dominant configuration dynamics -- three logical rules. These rules can also be intuitively argued from our kinetic-thermodynamic discrepancy design principle. 
As shown below, the next dominant configuration, given the present dominant configuration and control parameter $\lambda_i$, can be predicted by applying the three logical operations:
\begin{itemize}
    \item Pop-unfold rule: if more than i units are folded, unfold units from right to left until only i units are folded;
    \item Fold rule: if less than i units are folded, fold the rightmost unfolded unit.
    \item Exchange rule: if exact i units are folded, unfold the rightmost folded unit and fold the first unfolded unit on its left.
\end{itemize}
which is also illustrated in Fig.~\ref{fig:rules}a.

One can continue to apply the three rules above in accordance to the temporal control protocol to find the sequence of dominant configuration dynamics of a $N$-unit molecule. This dynamics resembles a series of binary bits evolving in time. If external control parameter $\lambda$ is fixed for all time, the dominant configuration eventually reaches the most probable configuration at the equilibrium distribution for the given control. The combination of these three rules provides an equivalent alternative representation of the dominant configuration evolution information described by the transition graphs shown in Fig.~\ref{fig:3unit}c.

\subsection{Complete State Controllability}
In this section, we demonstrate that our designed molecule can be steered into any desired configuration through appropriate control protocols. In other words, the designed molecule is completely controllable, where each one of the $2^N$ meta-stable configurations can --through external control protocol-- become the most probable configuration.

The complete controllability showcases the advantage of non-equilibrium processes over equilibrium or quasi-static processes. For instance, only $N+1$ configurations could become the dominant configuration under the $N+1$ control parameters $\lambda_i$. In contrast, different non-equilibrium driving protocols $\lambda(t)$ can be found to realize arbitrary dominant configuration.  

By utilizing the logical rules, we can come up with control protocols to (see Fig.~\ref{fig:rules}b) steer the dominant configuration into arbitrary desired configuration (e.g., ``0110110"). Specifically, the dominant configuration can be realized by sequentially preparing its two folded blocks starting from the left side (intermediate 0110000) to the right side (final 0110110). Here a folding block is a consecutive sequence of 1's. Our designed kinetics of the molecule allows a separation of timescale, and the operation time to prepare a block on the right hand side is much shorter than the evolution timescale of the blocks to its left. In Fig.~\ref{fig:rules}b lower panel, we showcase two strategies to prepare an individual folded block, and one can find that strategy B may be more time efficient than A.

\begin{figure*}[ht!]
    \centering
    \includegraphics{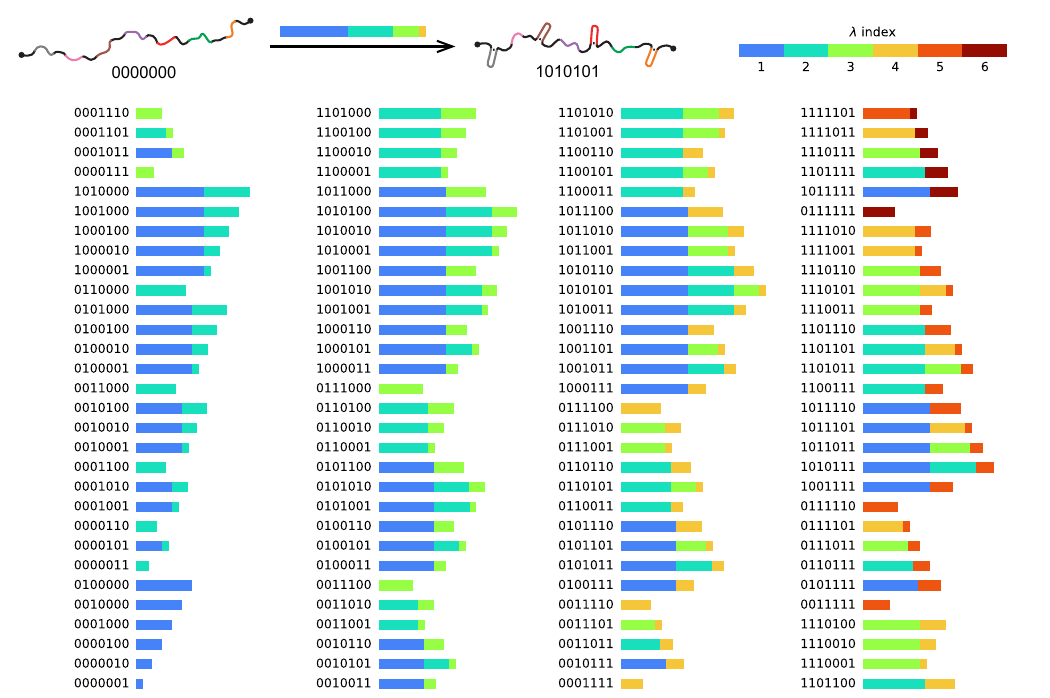}
    \caption{Demonstration of complete state controllability for a 7-unit molecular computer. Each color bar represents a unique non-equilibrium driving protocol to achieve one of the 128 possible dominant configurations. The color indicates the value of control $\lambda_i$, while the length of each bar represents the logarithm of the dwell time at that $\lambda_i$. Notice that, at equilibrium, only 8 configurations can become a dominant configuration. In contrast, the non-equilibrium driving protocols can steer to achieve any desired dominant configuration.}
    \label{fig:bar}
\end{figure*}

To illustrate the complete controllability, we enumerate the designed non-equilibrium protocols to achieve each one of the $2^N=128$ meta-stable configurations for a $N=7$ molecule as the dominant configuration. We initialize the molecule using $\lambda_0$, which stretches all units to an unfolded state, resulting in an initial dominant configuration of ``0000000". From this initial state, we can utilize either strategy A or B to design the steering protocols with different sequences of $\lambda_i$ with specific dwell times. The protocols constructed with strategy A are shown in Fig.~\ref{fig:bar}. Here the color indicates the value of $\lambda_i$ and the logarithm of the dwell time at each $\lambda_i$ is shown by the length of each color ribbons. 

The demonstrated complete state controllability of our molecular system has profound implications for potential applications. This property ensures that we can reliably set the molecule to any desired state, which is crucial for using it as a programmable device. In the context of information processing, this means we can use the molecule as a rewritable memory or a reconfigurable logic gate. For sensing applications, complete state controllability allows us to reset the sensor or adjust its sensitivity by preparing specific initial states. In more complex applications like molecular robotics \cite{thubagere2017cargo,hagiya2014molecular,liu2019biomedical} or adaptive materials \cite{rybtchinski2011adaptive,blum2015stimuli}, this property would enable the system to be reprogrammed or adapted to new tasks on the fly. Essentially, complete state controllability transforms a single molecule into a versatile, multi-purpose nanoscale device.

\section{Discussion}
\subsection{Molecular automaton}

In computer science, an automaton can carry out language recognition. It consists of a set of states, an input alphabet, and rules of state transition. Given an initial state and a sequence of input string of symbols from the alphabets, the machine's state evolve according to the rules into a final state; if the final state belongs to the \emph{accept set}, then the machine accepts that the input string, otherwise the machine rejects the input string \cite{Moore2011-pa,hopcroft2001introduction}. 

The proposed molecule for temporal pattern recognition is similar to the automaton. To illustrate the analogy, the $2^N$ configurations of the molecule corresponds to the automaton's set of states; the $N+1$ controlled end-to-end distances ${\lambda_i}$'s correspond to the input alphabet; the temporal protocol $\lambda(t)$ corresponds to the automaton's input sequence. Additionally, the logical rules of dominant configuration evolution in Fig.~\ref{fig:rules}a fully describes the state transition rules of the automaton: given a current dominant configuration and a $\lambda_i$, the next dominant configuration is determined by one of the 3 logical rules. Ultimately, if we select one or a few configurations as ``accept states" (e.g. an active configurations that can emit signal to the downstream sensory network), then the molecule can be seen as an functional automaton to report temporal patterns of its mechanical input to downstream sensory networks.

This work bridges the gap between the information processing devices that faithfully perform computational tasks and soft macromolecules whose configuration evolves in complex and stochastic manner.  
The apparent difficulty of implementing deterministic automaton from computer science into the stochastic configurational transitions is resolved at the level of the dominant configuration.  By focusing on the dominant configuration of the molecule, the designed molecule is shown to deterministically obey three logic rules (see Fig.~\ref{fig:rules}a). As a result, one can harness the rich, non-equilibrium behavior of a single molecule for faithfully performing sophisticated information processing tasks, opening up the possibility of molecular-scale computation and temporal pattern recognition.

In the following we discuss two different perspectives in the application of such molecular automaton -- sensory pattern recognition and transient memory device. 

\subsection{Sensory Pattern Recognition}

The first application of the molecular automaton is to sense temporal patterns of environmental signals and transduce information to the downstream sensory network. 
For example, one can construct a \emph{molecular combination lock}, which can be unlocked (activated) only by a correct sequence of input signal $\lambda(t)$. Confider this molecule can only achieve a certain dominant configuration via a specific sequence of input signal, i.e., the password sequence $\lambda(t)$. If the configuration is able to catalyze the downstream signaling reaction or trigger florescent activities, then this molecule becomes a combination lock that is only switched on after the correct temporal pattern signal is provided. This example demonstrate that, given a complex energy landscape with numerous states kinetic frustration, a single molecule can perform rather complicated information sensory and processing tasks. This capability is particularly relevant in fields such as biochemical sensing\cite{changeux2005allosteric}, drug delivery \cite{douglas2012logic,zhao2012dna,liu2021multifunctional}, molecular machines \cite{henzler2007dynamic,bath2007dna}, and self-assembly systems \cite{evans2024pattern,yin2008programming}, where the recognition and processing of time-varying signals are crucial.

\subsection{Transient Memory Device}

The hysteric molecule achieving non-equilibrium states in accordance with the driving protocol can be considered a transient memory device. Unlike covalent-bound based memory molecules that offer stable, long-term storage (e.g., base-pair sequence stored in DNA), our system provides a dynamic, easily rewritable memory with shorter retention times. 

In this device, information is stored in the format of binary bit streams (e.g., ``0010101''), represented by the polymer's dominant configuration. In other words, the molecule with $N$ units constitutes a short-term memory device with memory capacity of $N$ bits.

For the designed molecule its non-equilibrium memory far exceeds the equilibrium memory capacity. The accessibility of the memory state space is a consequence of non-equilibrium driving forces. At thermal equilibrium, there are only $N+1$ possible dominant configurations. In contrast, the designed molecule is shown to process $2^N$ dominant configurations. In other words, the equilibrium memory capacity of the molecule is $\log_2(N+1)$ bits, whereas the non-equilibrium memory capacity of the molecule is $N=\log_2(2^N)$ bits.

The information writing and erasing protocols can be determined by the logical evolution rules as shown in Fig.~\ref{fig:rules}a. 
The information erasing operation, which resets the dominant configuration into ``0,0,0,...,0" regardless of the molecule's previous states, can be realized by a simple mechanical pulling with $\lambda_0$. After information eraser, one can write any desired bit string with the protocols determined by the automaton's rule. A complete set of information writing protocols to achieve all 7-bit streams are demonstrated in Fig.~\ref{fig:bar}. 

The performance of the memory can be characterized by \emph{memory retention time}, \emph{information writing speed}, and its \emph{operational energy cost}. Unlike DNA, which stores information in covalent bonds\cite{goldman2013towards,tabatabaei2015rewritable,erlich2017dna,organick2018random}, this molecular memory device only transiently retains data in its metastable configuration. Consequently, memory retention time is a crucial performance characteristic. Due to the designed time-scale separation, bit strings with "1" near the right-hand side have shorter retention times than those without. The practical usefulness of the memory, as a result, is dictated by the comparison between the time it takes to write information and the memory retention time. As shown in Fig.~\ref{fig:rules}b, there are tow strategies in deciding the writing protocol. 
In strategy A, the control length is only relaxed in steps to form blocks one after another, and there is no pulling. 
In comparison, when forming a 1 block, strategy B let the molecule relax at the shortest control length so more transitions happen and it folds much faster, then followed by a pulling to trim the extra 1s on the right.
Strategy B achieves a shorter information writing time by first allowing many units to fold followed by a sudden mechanical pulling to pop open the undesired folded units.
As predicted by the Landauer principle, for a memory device operating at temperature $T$, it takes at least $k_BT \ln 2$ joules of energy to erase 1 bit of information \cite{landauer1961irreversibility,bennett2003notes,lu2014engineering}, here the memory device's operational energy expenditure can vary from strategy to strategy. An obvious tradeoff between the information writing speed and the energy cost can be argued as follows. In strategy B, the writing protocol takes less time but involves additional mechanical pulling operations to pop open folded units, thus resulting in higher energy cost in the memory erasing and writing cycle (see SI).

\subsection{Implementation by DNA or other realistic systems}

While the molecular automaton model provides a powerful conceptual framework, several challenges must be addressed for practical implementation. These include scalability issues for larger systems, potential error rates due to thermal fluctuations, development of efficient read-out mechanisms, considerations of operational time scales, and energy requirements. Below we sketch possible approaches to address these challenges toward experimental realization of the proposed smart molecule.

The theoretical framework presented here could be experimentally realized using nucleic acid engineering, particularly DNA nanotechnology\cite{manosas2005thermodynamic,collin2005verification,hummer2001free,dudko2008theory,soloveichik2010dna}. Foldable units could be implemented as DNA hairpins, with stem length and sequence composition precisely tuned to achieve desired folding energies and kinetics \cite{zadeh2011nupack,soloveichik2010dna,zhang2009control,rothemund2006folding,torelli2014dna,sacca2012dna,douglas2009self}. More complex three-dimensional structures with controlled folding pathways could be created using DNA origami techniques \cite{rothemund2006folding,torelli2014dna,sacca2012dna,douglas2009self}. Furthermore, folding around core molecules such as ligands, proteins, or inorganic nanoparticles could dramatically alter the energetic landscape, acting as nucleation sites or conformation stabilizers. These approaches, individually or in combination, offer a promising avenue for implementing our proposed smart polymer design, potentially enabling the creation of molecular-scale computers and sensors with programmable responses to complex environmental signals.

The proposed molecular automaton opens avenues for diverse practical applications. By associating molecular configurations with gene expression, we could create gene switches responsive to complex temporal signals, enabling pattern recognition in cellular environments. Alternatively, linking configurations to fluorescence \cite{ishii1999fluorescence,kruger2013structural,lippincott2001studying} could yield environmental sensors that emit light upon detecting specific temporal patterns. A limitation of the proof-of-principle model is the exponential separation of relaxation timescales for each foldable unit. This constraint, however, can be overcome by employing dissipative dynamics in the folding-unfolding transitions, such as through the application of active driving forces. By introducing dissipative energy into the smart molecule's transitions, a wider range of dynamics can be achieved, potentially enabling the decoding of more complex temporal patterns. This enhancement could significantly expand the capabilities of molecular-scale computation and sensing in diverse fields including biophysics and material sciences.

\subsection{Future Directions}
In conclusion, this work presents a novel approach to molecular-scale computation and information processing, bridging the gap between deterministic computational devices and stochastic molecular systems. By leveraging the concept of thermal-kinetic frustration in a single molecule, we have demonstrated the potential for creating molecular automata capable of complex temporal pattern recognition and transient information storage. The proposed system offers unique advantages in terms of programmability, adaptability, and non-equilibrium operation, opening up new possibilities in fields ranging from biosensing and drug delivery to adaptive materials and molecular robotics. While challenges remain, particularly in experimental implementation and scaling, the framework presented here provides a solid foundation for future research. Key directions for future work include: (1) experimental realization of the proposed molecular automaton using DNA nanotechnology, (2) exploration of dissipative dynamics to overcome relaxation timescale limitations, (3) development of efficient read-out mechanisms for practical applications, and (4) investigation of more complex computational paradigms that could be implemented using this molecular framework. As we continue to push the boundaries of molecular-scale computation, systems like the one proposed here may play a crucial role in developing the next generation of smart, responsive materials and devices.

\section{Methods}
\subsection{Unit-wise design of Ising-type molecule}
The design of our model molecule is an Ising-model-like linear polymer chain that comprises $N$ binary-state foldable segments connected by inert segments. Each foldable segment (or unit) can hop between folded state (``1'') and unfolded state (``0''). An isolated segment (e.g., the i-th unit) can fold and unfold according to its binary-state energy landscape, where the unfolded state energy is $0$, folded state energy is $\Delta E_i$, and transition barrier is $B_i$.
The polymer thus has $2^N$ possible configurations, each denoted by a $N$-bit string, $\bf{s}$.

To achieve the kinetic-thermodynamic discrepancy in the $N$-unit Ising-type polymer, we utilize a mismatching trend of the unit-wise folded energies and the folding activation barriers. The design is illustrated for a $N=7$-unit polymer. The folded energies $E_{f,i}$ of units $i$ from 1 to 7 increases with respect to $i$, taking values as [-7, -6, -5, -4, -3, -2, -1] in reduced units. In an opposite trend, the fold-unfold barrier heights $B_i$ for unit $i$ decreases with respect to $i$, taking values of [7, 6, 5, 4, 3, 2, 1] in reduced units. We assume that the unfolded energies are all equal to $E_{u,i}=0$ for every unit.

\subsection{Mechanical control of linear molecule}
The configuration of the polymer defines an effective contour length, $l({\bf{s}})$, which is the maximum end-to-end distance that the polymer can sustain without unfolding any folded units of the configuration. 
By assuming that the folded state and the unfolded state has a length difference $\Delta l$ for all segments, the effective contour length for any configuration $l({\bf{s}})$ is
\begin{equation}
l({\bf{s}}) = l_\text{max} - N_f({\bf{s}})\cdot \Delta l
\end{equation}
where $l_\text{max}$ is the contour length of the fully unfolded configuration $\bs=(0,0,0,\cdots)$, and $N_f({\bf{s}})\equiv \sum \bf{s}$ is the number of folded segments (state 1's) of the configuration $\bf{s}$. For such a polymer, $l({\bf{s}})$ can assume $N+1$ possible values. Here we use $l_\text{max}=20$ and $\Delta l=2$.

The polymer is mechanical controlled (pulling and relaxation) by a temporal protocol of the distance between the two ends of the polymer, denoted by $\lambda(t)$. 
As a result, the polymer undergo configuration transitions that are affected by the intrinsic folding and unfolding energies of each unit, the external controlled distance $\lambda$ at the given time, and the contour length corresponding to the transient configuration.

Consider an initial configuration $\bs_0$, if the end-to-end distance is set to $\lambda$, two types of transitions could occur: the popping transition and the stochastic transitions. (1) If the initially configuration is not mechanically compatible with the end-to-end distance, i.e., $l(\bs_0)<\lambda$, then the weakest foldable unit(s) popping unfold, until the polymer contour length is larger than the given $\lambda$. The weakness of the unit $i$ is defined by the activation barrier $B_i-\Delta E_i$ of the unfolding event. 
(2) If the initial configuration is compatible with the end-to-end distance, i.e., $l(\bs_0)>\lambda$, thermal fluctuations allow each unit to fold or unfold, as long as the effective contour length is always compatible with the given distance ($l(\bs)>\lambda$). The configuration transition rates depend on both the configurational energy and the polymer's chain entropy $S(l,\lambda)$. Thus, the folding and unfolding kinetics of each unit is implicitly dependent on the states of the others and on $\lambda$. 

For a given polymer with $N$ units, the contour length $l(\bs)$ can take $N+1$ different values $l_i=l_{\text{max}}-i\cdot \Delta l$, where $i=0,1,\cdots,N$ denotes the number of folded units. To capture the mechanical driving protocol of the molecule, we assume that at any time, $\lambda(t)$ can only assume $N+1$ discredited values $\lambda_i = l_N-\Delta l/2$. Under this assumption, given the distance $\lambda_i$, the polymer can only allow for with up to $i$ folded units.
As a result, the mechanical control protocol is denoted by $\lambda(t)$ that is piece-wise-constant function of time.

\subsection{Stochastic dynamics of controlled molecule}
We use a vector $\vec p$ to describe the probabilities of all $2^N$ configurations the molecule.
We separate the time scale of rapid pulling caused by instant changing of $\lambda$ and the slow stochastic transitions.
At the instantaneous time of switching $\lambda$, the molecule maybe forced to a rapid unfolding(s), which can be characterized by an instant transition matrix $W^Q$, based on the unfolding barrier of each units. 
At any constant $\lambda$, the slow stochastic transition dynamics can be described by a slow transition matrix $W^R$, which can be solved by the
follows the master equation:
\begin{equation}
    \frac{\mathrm{d}\Vec{p}}{\mathrm{d}t} = R_\lambda \Vec{p}
    \label{eqn:relaxation}  
\end{equation}
where the transition probability rate matrix $R_\lambda$ characterizes the transition rates for each folding and unfolding dynamics due to thermal fluctuations. 
For a time duration $\tau$, the slow transition matrix can be expressed as
\begin{equation}
        W^R_{\lambda}(\tau)=\exp(R_{\lambda} \tau)
\end{equation}
Thus, the propagator for a sudden switch of control length to $\lambda$ and a period of constant control$\tau$ can be expressed by the product of the fast and slow probability transition matrix $W^R_{\lambda}(\tau) \cdot W^Q$.

The transition matrix of the slow dynamics $R(\lambda)$ under a given control $\lambda$ can be determined from the free energy analysis of the polymer. 
We assume that the entropy of the polymer given a configuration (contour length $l$) and a end-to-end distance $\lambda$ can be calculated as an ideal chain with contour length $l$ and a fixed end-to-end distance $\lambda$ as $S = \log(l^2 - \lambda^2)$ \cite{winkler1992finite}. 
Then the free energy of a configuration can be found by $F = E - T S$, where $T$ is the temperature of the bath.
Here we set $T$ to 0.6.
The slow stochastic transition rate matrix are constructed using $R_{\lambda,y,x} = \exp\{-\beta (B_{yx}-F_x)\}$, where configuration $x$ and $y$ are two mechanically allowed configurations given $\lambda$, and the two configurations are only different by folding/unfolding one unit.
$\beta$ is the inverse temperature, $\beta = 1/T$.
The folding/unfolding activation energy barriers $B_{yx}$ between configurations $x$ and $y$ are determined by adding a designated barrier value from the unfolded state.
The diagonal elements $R_{\lambda,x,x}$ are chosen such that each column sums to $0$.

The probability propagator $W^Q_{\lambda_i}$ corresponding to the fast forced open of units can also be determined step-by-step. Specifically, under a pulling control $\lambda_i$, the propagator maps all probabilities of mechanically forbidden states to the states with $i$ folded units. This propagation can be considered as a sequence of steps, each taking infinitely short time:
\begin{equation}
    W^Q_{\lambda_i} = W_i \cdot W_{i+1}  \cdots W_{N-2}\cdot W_{N-1}
\end{equation}
where $W_j$ represents the fast pulling step resulting from the end-to-end-distance reduction from $\lambda_{j+1}$ to $\lambda_j$. For each step, its step-wise propagator matrix only takes non-zero off-diagonal elements for transitions from a $(j+1)$-unit-folded state $x$ to a $j$-unit-folded unit $y$:
\begin{equation}
    W_{j,y,x} = \exp\{-\beta (B_{xy}-F_y)\}/Z
\end{equation}
where $Z$ is the normalization factor to ensure each column sums up to 1. For the matrix columns not involved in the transitions described above, its diagonal element is chosen to be 1, while the rest elements of the column equal zero.

We assume that the temporal control protocol $\lambda(t)$ takes a piece-wise-constant form of $\lambda(t)=\lambda_{(k)}, \forall~  t_{k-1}<t < t_{k} $ which assumes each constant value $\lambda_{(k)}$ lasts for duration $\tau_k$, and $t_k$ denotes the ending time of the the $k$-th step, $t_k = \tau_1+ \cdots +\tau_k$. In this notation, the initial time is set to $t_0 = 0$. 

Under this piece-wise constant control protocol, the total propagator $\mathcal T$ for a $K$-step control protocol is simply the product of propagators corresponding to each time period:
\begin{equation}
    \mathcal T  = \prod_{k=1}^{K} (W^R_{\lambda_{(k)}}(\tau_k) \cdot W_{\lambda_{(k)}}^Q )
    \label{eqn:solution}
\end{equation}
Eqs.~\ref{eqn:solution} can be applied to arbitrary protocols of sudden increases and decreases of end-to-end distance (arbitrary piece-wise constant function $\lambda(t)$).

\begin{acknowledgments}
This work is supported by the University start-up fund at UNC-Chapel Hill. We acknowledge the fund from the National Science Foundation Grant DMR-2145256.
\end{acknowledgments}

\bibliography{apssamp}

\end{document}